\documentclass[fleqn,usenatbib]{mnras}

\usepackage{newtxtext,newtxmath}

\usepackage[T1]{fontenc}

\DeclareRobustCommand{\VAN}[3]{#2}
\let\VANthebibliography\thebibliography
\def\thebibliography{\DeclareRobustCommand{\VAN}[3]{##3}\VANthebibliography}

\usepackage{graphicx}	
\usepackage{amsmath}	

\title[Spins of galaxies in the Local Volume] {Orientation of the spins of galaxies in the Local volume}

\author[I.D. Karachentsev \& V.D. Zozulia]{
I.D. Karachentsev ,$^{1}$\thanks{E-mail: idkarach@gmail.com}
V.D. Zozulia$^{2}$\\
\\
$^{1}$Special Astrophysical Observatory of the Russian Academy of Sciences, Nizhnĳ Arkhyz, Karachay-Cherkessia 369167, Russia\\
$^{2}$St. Petersburg State University,
Universitetskij pr.~28, 198504 St. Petersburg, Stary Peterhof, Russia
}

\date{Accepted XXX. Received YYY; in original form ZZZ}

\pubyear{2023}

\begin{document}
\label{firstpage}
\pagerange{\pageref{firstpage}--\pageref{lastpage}}
\maketitle

\begin{abstract}
We estimated the angular momentum, $J$, of $720$ galaxies in the Local Volume
with distances $r < 12$ Mpc. The distribution of the average angular momentum along the Hubble sequence has a maximum at the morphological type $T= 4$ (Sbc), while the dispersion of the $J$-values for galaxies is minimal. Among
the Local Volume population, 27 elite spiral galaxies stand out, with an angular momentum greater than 0.15 of the Milky Way, $J > 0.15 J_{MW}$, making the main contribution ($ > 90\%$) to
the total angular momentum of galaxies in the considered volume. Using observational data on the kinematics and structure of these galaxies, we determined the direction of their spins.
 We present the first map of the distribution of the spins of 27 nearby massive spiral galaxies in the sky and note that their pattern does not exhibit significant alignment with respect to the Local Sheet plane. The relationship between the magnitude of the
angular momentum and stellar mass of the local galaxies is well represented by a power law with
an exponent of ($5/3$) over an interval of $6$ orders of magnitude of the mass of
galaxies.
\end{abstract}




\section{Introduction}

According to modern cosmological concepts, the angular momentum (spin) is a fundamental property of a galaxy and contains information about the origin and evolution of stellar systems. Spins of protogalaxies are generated by a tidal torque in a local field during the formation of the large-scale structure of the Universe \citep{Peebles_1969, Doroshkevich_1970, White_1984}. In this paradigm, galactic spins show a mutual alignment which remain preserved for many billions of years and are broken during episodic mergers of galaxies. Spins of galaxies located in the plane of the cosmic “pancake” have a characteristic feature. They are more likely to be oriented along the pancake plane. In other words, the disk planes of spiral galaxies tend to be perpendicular to the plane of the pancake.

\citet{Codis_et_al_2018} and  \citet{Ganeshaiah_et_al_2019, Ganeshaiah_et_al_2021} performed   the cosmological 
hydrodynamical simulation and found that the spin of low-mass galaxies tends to be 
aligned with the filaments of the cosmic web and to lie within the plane of the walls while 
more massive galaxies tend to have a spin perpendicular to the axis of the filaments and to 
the walls. \citet{Kraljic_et_al_2020}  studied  the spin alignment of galaxies and haloes with respect 
to filaments and walls of the cosmic web,  using the SIMBA simulation and also concluded that
massive haloes’ spins are oriented perpendicularly to their closest filament’s axis and walls, while
low-mass haloes tend to have their spins parallel to filaments and in the plane of walls.

Many authors have tested theoretical predictions about the alignment and orientation of the galactic spins relative to the elements of the large-scale structure
\citep{Kapranidis_Sullivan_1983a, MacGillivray_Dodd_1985a, Lee_Pen_2002, Tempel_Libeskind_2013, Zhang_Yang_Wang_Wang_Luo_Mo_van_den_Bosch_2015, Antipova_Makarov_Bizyaev_2021}. 
Some of them used edge-on or face-on spiral galaxies. It allowed to fix one of two mutually opposite spin directions. As a result, weak signs of anisotropy in the orientation of galaxy spins have been found \citep{Shamir_2016, Antipova_Makarov_Bizyaev_2021}.

The observational analysis of spin orientations in some articles refer to the volume of the Local Supercluster of galaxies. It is a flat structure approximately 40 Mpc in size, centred on the Virgo cluster and observed edge-on. Despite there being a large number of spiral galaxies in this volume ($N \approx 10^3$), data on the orientation of the angular momenta of these galaxies are still extremely scarce. \citet{Navarro_Abadi_Steinmetz_2004}  examined the distribution of 30 spiral edge-on galaxies from the PGC catalogue \citep{Paturel_1997} by position angle of the major axis, and found signs of the expected preferred location of the galactic spins in the plane of the Local Supercluster. However, this result needs to be confirmed because it has a low statistical significance.

\citet{Tempel_Stoica_2013}  and \citet{Tempel_Libeskind_2013} used the Sloan Digital Sky Survey to study
the connection between the spin axes of galaxies and the orientation of their host filaments.
The authors  found evidence that the spin axes of bright spiral galaxies have a weak tendency to be
aligned parallel to filaments. For E/S0 galaxies, they obtained a statistically significant result
that their spin axes are aligned preferentially perpendicular to the host filaments. \citet{Blue_Bird_et_al_2020}
, \citet{Welker_et_al_2020},\citet{Kraljic_et_al_2021}, and \citet{Barsanti_et_al_2022} investigated
the alignments of galaxy spin axes with respect to cosmic web filaments as a function of various 
properties of the galaxies and their constituent bulges and discs, using different samples of
galaxies, and found that  galaxies with lower bulge masses tend to have their spins parallel to the 
closest filament, while galaxies with higher bulge masses are more perpendicularly aligned. At
total, the measured alignment signal is found to be weak, but statistically significant.

Obviously, the approach that uses the most complete information about the orientation of the galactic spins in the closest volume of space deserves attention. The first such attempt was made by \citet{Karachentsev_1989}, where the spin directions were determined for 20 disk galaxies in the Local Group and the neighbouring group around the M81 galaxy. Over the last 3 decades, new data on distances, structure, and kinematics of nearby galaxies have appeared. It allows extending this approach to the population of the entire Local Volume (LV) enclosed in a sphere with a radius of 12 Mpc around the Milky Way (MW).

A summary of various observational data on LV galaxies is presented in Updated Nearby Galaxy Catalog (UNGC), \citep{Karachentsev_Makarov_Kaisina_2013}. Its latest updated version (www.sao.ru/lv/lvgdb) contains about $1300$ galaxies. A significant part of them ($N \approx 500$) have accurate distance estimates obtained with the Hubble Space Telescope. The spatial structure of LV and its neighbourhood was described in detail in 
\citet{Courtois_Pomarede_Tully_Hoffman_Courtois_2013}. The main feature of the LV is a flat structure the so-called Local Sheet \citep{Tully_2008} ,  where the majority of the LV population is concentrated. The plane of the Local Sheet coincides with the plane (equator) of the Local Supercluster within a few degrees. According to \citet{Karachentsev_Makarov_Kaisina_2013}, the Local Sheet is approximately the shape of a human palm, as it is a triaxial ellipsoid, the major, middle, and minor axes of which are 10 Mpc, 7 Mpc and 2 Mpc, respectively. The centre of this ellipsoid is shifted relative to the MW by about 3 Mpc in the direction of the Leo-Virgo constellations. The distribution of galaxies outside the Local Sheet is asymmetrical: a vast almost empty region (Local void) is located to the north along the supergalactic axis $SGZ> 0$, the neighbouring flat structure the Leo Spur \citep{Tully_Fisher_1987}, is located to the south at $SGZ< 0$.

\begin{figure*}	\includegraphics[width=\textwidth]{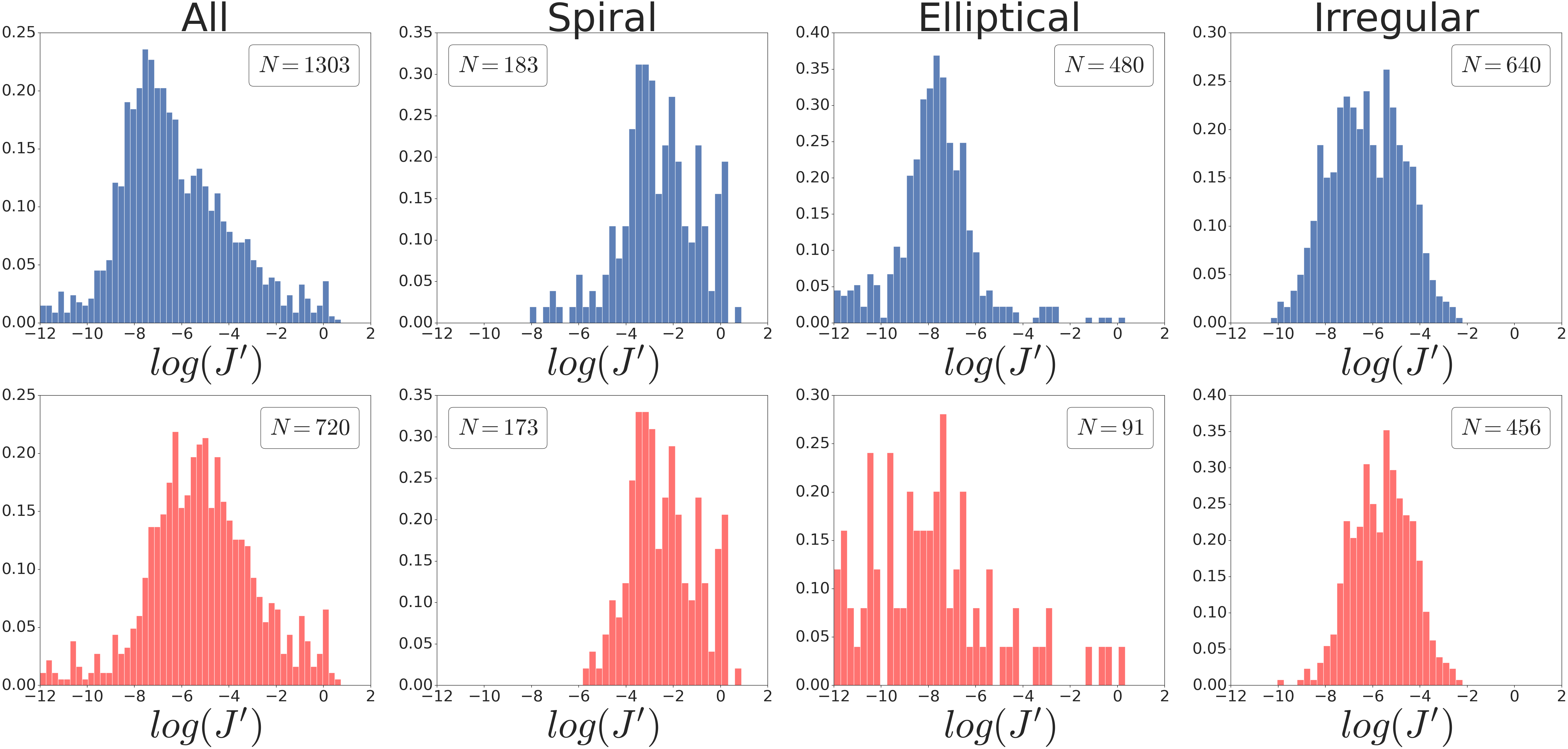}
\caption{Distribution of galaxies of the Local Volume of different morphological types according to the magnitude of the angular momentum. \textit{Below}, distributions of galaxies with known estimates of the rotation amplitude $V_m$. \textit{Above}: the distribution with allowance for galaxies without $V_m$ estimates, to which a rotation amplitude of $1$ km s$^{-1}$ is assigned.}
    \label{fig:hist_J_LV_all}
\end{figure*}

\begin{figure}
\centering
\includegraphics[width=\columnwidth]{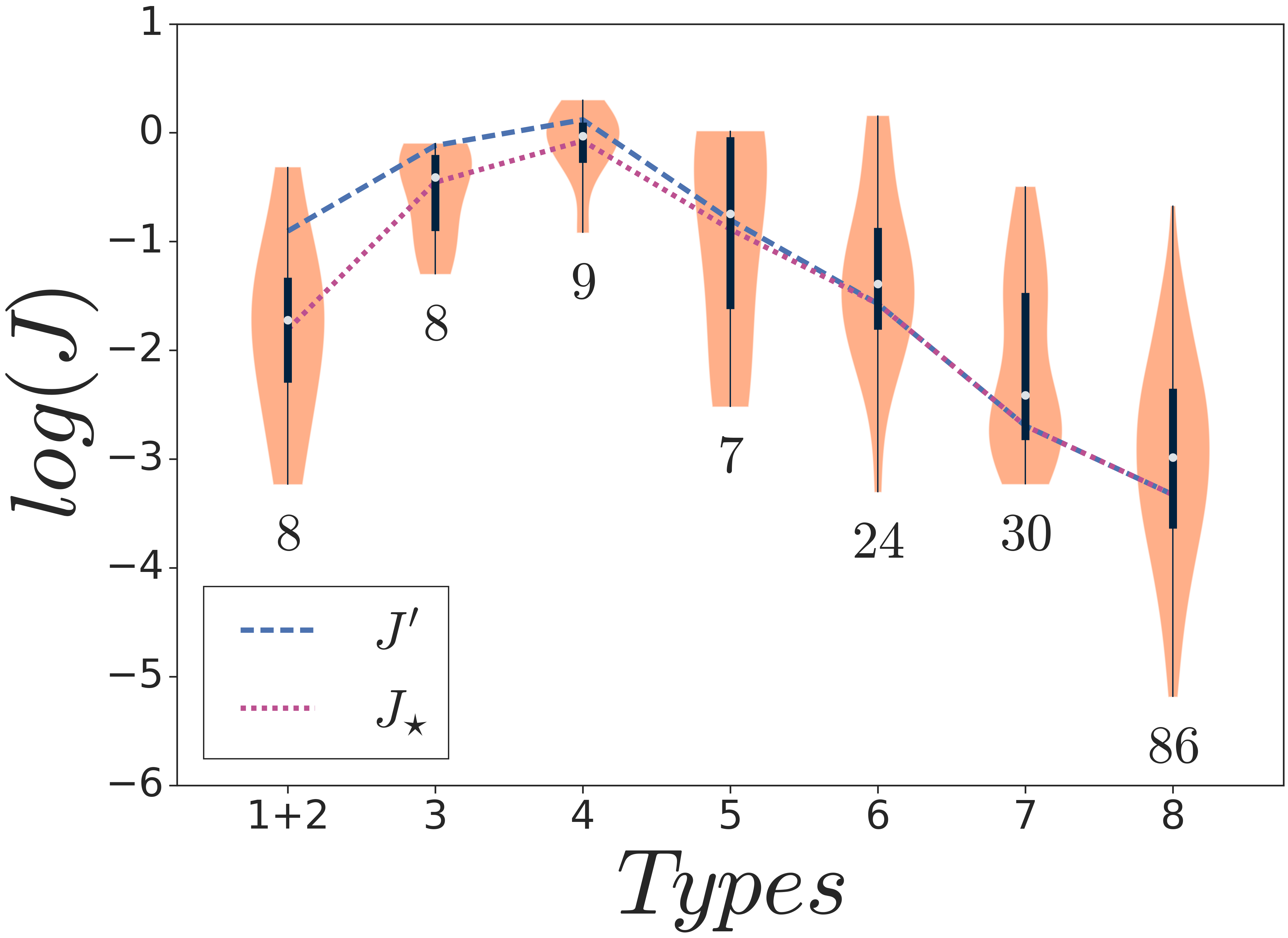}
    \caption{Dependence of the median angular momentum on the morphological type of spiral galaxies.}
    \label{fig:violin}
\end{figure}

\section{Angular momentum of rotation of the galaxies in the Local
Volume}

In order to calculate the total angular momentum of a galaxy's rotation, it is necessary to have detailed data on its structure and kinematics, in particular, the distribution of stellar density along the radius and the rotation curve to the farthest boundary of the galaxy. Such data are known so far only for a few galaxies. Therefore, below, we use an approximate estimate of the angular momentum

\begin{equation}
    J' = M_\star  \times V_m \times R_{26},
	\label{eq:J'}
\end{equation}

here $M_\star$ --- stellar mass of the galaxy, $V_m$ --- maximum amplitude of its rotation and $R_{26} = A_{26}/2$ --- the Holmberg radius of the galaxy measured on isophote $26.5^m$ per square arcsecond through a B filter.  $M_\star$ and $A_{26}$ are presented for all LV galaxies in the UNGC catalogue. The amplitudes of rotation, $V_m$, are known only for 55\% of this sample. The stellar mass was estimated from the luminosity of the galaxy in the K-band. For some dwarf galaxies with very low surface brightness, we used their effective radius, within which half of total luminosity is concentrated, instead of $R_{26}$. When the galaxy's $V_m$ was determined (from the optical rotation curve or from the width of the HI 21-cm line), a correction was introduced for the inclination of the galaxy and turbulent motions in it. Details regarding the definition of these values are described in the UNGC.

The values of the angular momentum $J'$ were calculated using expression (\ref{eq:J'}) for all 720 LV galaxies with estimates of $V_m$. We normalized them to the value of the angular momentum of the MW. The following parameters were chosen for the Galaxy \citep{Bland-Hawthorn_Gerhard_2016} : $M_\star = 5\cdot10^{10}\,M_\odot,\; M_{bulge} = 1.5 \cdot 10^{10}\,M_\odot, \;M_{gas} = 1.3\cdot 10^{10}\,M_\odot, V_m = 238$ km s$^{-1}, A_{26} = 2 \times R_{26} = 30$ kpc. The last value was estimated in a different way. The first method assumed an exponential density distribution, a fixed disk mass $3.5\cdot10^{10}\,M_\odot$ and the mass-to-light ratio in the B-band $\Upsilon_B = 1.2 \,M_\odot/L_\odot$, then $R_{26}$ varies from $15.2$ to $17.2$ kpc when changing the radial scale from $2.5$ to $3$ kpc. The second way used the empirical dependence between $R_{26}$ and the total luminosity $L_K$ for galaxies of the morphological type $T=4$, which gave the estimate for the MW $A_{26} = 28\pm2$ kpc.

The distribution of relative number of LV galaxies in terms of the normalized angular momentum is shown in the bottom panels of Fig.~\ref{fig:hist_J_LV_all}. The left panel shows the distribution for galaxies of all morphological types, $T$. 
The three right panels demonstrate separately distributions for spiral galaxies ($T = 1, ...8$), elliptical ($T \leqslant 0$), and irregular ($T = 9$ and $10$). As expected, the angular momentum distributions differ significantly for galaxies of different types. Their distribution maxima differ by several orders of magnitude. It should be noted that the galactic angular momentum distribution becomes unreliable at values of  $J' /J_{MW} \leqslant 10^{-5}$, since about $45\%$ of LV galaxies do not have $V_m$ estimates. We assigned to these galaxies (mainly dwarfs) the formal value of the rotation amplitude $V_m \approx 1$ km/s. The top panel in Fig.~\ref{fig:hist_J_LV_all} shows distributions augmented by dwarf galaxies. The number of galaxies in each sample is indicated in the upper right corner of each panel. In the case of spiral galaxies, the increase in objects without $V_m$ was insignificant, but the number of early-type galaxies increases almost $5$ times.

We considered $10$ LV spiral galaxies  without estimates of $V_m$ and found that they all belong to the late types Scd, Sd ($T = 7,\,8$) with a low median luminosity $\log(L_K/L_\odot) = 7.82$. Counting or ignoring these galaxies has little effect on the overall shape of the distribution  $N(J')$. It should be noted that among early-type galaxies ($T \leqslant 0$) there are $9$ cases with estimates of $J' /J_{MW} > 10^{-3}$. They are presented in Table~\ref{tab:Table1}. Its columns contain: (1) the name of the galaxy; (2) the angular moment of rotation; (3) the distance in Mpc; (4, 5)—morphological type according to de Vaucouleurs digital scale from the UNGC catalogue and also from the HyperLEDA database \citep{Makarov_Prugniel_Terekhova_Courtois_Vauglin_2014}. The rotation amplitudes of these galaxies are estimated from the motions of gas, the kinematics of which differs significantly from the kinematics of the stellar population. This means that the values of their angular momenta might be overestimated.

\begin{table}
   \caption{Early type ($T < 1$) galaxies 
                 in the LV with $\log(J') > -3.0$}
\begin{tabular}{lrrrr}
 \hline
 Name         &  $\log(J'/J_{MW})$ &   $r$ (Mpc)  &   $T_{\rm UNGC}$  &   $T_{\rm LEDA}$\\   
\hline
 NGC 5128     &   0.26   & 3.68  &  -2   &   -2.1   \\    
 NGC 3115     &  -0.40   & 9.68  &  -1    &  -2.9   \\     
 NGC 5195     &  -0.73   & 7.66  &   0   &     0.6  \\      
 NGC 6684     &  -1.27   & 8.70  &   0   &   -1.9    \\    
 NGC 3379     &  -2.58   & 11.32 &   -3  &    -4.8   \\     
 NGC 2784     &  -2.67   & 9.82  &  -2   &   -2.1   \\      
 Maffei 1     &     -2.71& 5.60  & -3    &   -3.0   \\     
 NGC 855      &   -2.82 &  9.73 &   0   &    -4.8   \\     
 NGC 404      &   -2.99  &  2.98 &  -1   &    -2.8  \\      
 \hline
 \end{tabular}

\label{tab:Table1}
 \end{table}

The bulges of spiral galaxies, like the population of elliptical galaxies, have a low rotation speed. The bulge/disk mass ratio depends on the morphological type of the galaxy, decreasing with increasing T. The relative fraction of the disk mass $f_{\text{T}} = M_{\text{disk}} /M_{\text{total}} $ for different types of spirals T can be represented by the approximate relation:

  \begin{equation}
   \begin{array}{ll}
  f_T=0.18\times T-0.08 & {\rm when} \,\, T= [1,2.\ldots6]\\
  f_T=0 & {\rm when} \,\,T<1\\
  f_T=1 & {\rm when}\,\, T>6.  
  \end{array} 
 	\label{eq:f_type} 
  \end{equation}

We neglect the angular momentum of the bulge and adopt the expression for the angular momentum of the galaxy
  \begin{equation}
  J_\star=f_T\times M_\star\times V_m\times R_{26}=0.3\times f_T\times L_K\times V_m\times A_{26}.
   	\label{eq:J_star} 
  \end{equation}

Here, following \citet{Lelli_McGaugh_Schombert_2016}, we assume that the stellar mass of a galaxy is proportional to its K-band luminosity, $M_\star = \Upsilon_K \times L_K$, with the coefficient $\Upsilon_K =0.6$ in units of the solar mass and luminosity. The used values $L_K, V_m, A_{26}$  and $T$ are taken from the latest version of the LV database, www.sao.ru/lv/lvgdb.

The gas component, along with the stellar component, makes a significant contribution to the total angular momentum of rotation in late-type galaxies. Thus, we introduce the concept of total (baryonic) angular momentum and define it as

  \begin{equation}
  J=(0.6\times f_T\times L_K+1.4M_{\rm HI})\times V_m\times (A_{26}/2)
     	\label{eq:J_all} 
  \end{equation}

where $M_{\rm HI}$ is the mass of hydrogen in the galaxy, and the factor $1.4$ takes into account the contribution of heavier elements to the gas mass. Hydrogen mass estimates for the galaxies are also contained in the Local Volume database.

A violin plot in Fig.~\ref{fig:violin} demonstrates the distribution of angular momentum for spiral galaxies of different morphological types. The white dots show the median values, the thin black lines show the range of values, and the bold lines represent the $25\%$ and $75\%$ quantiles. As we can see, the median angular momentum of spiral galaxies grows along the Hubble sequence from types $T = 1,\, 2$, 3 (Sa, Sab, Sb), reaching a maximum at $T = 4$ (Sbc), and then monotonically decreases towards $T = 5$ (Sc), $T = 6$ (Scd), $T = 7,\,8$ (Sd, Sm), and $T = 9$ (Im). The smallest interquartile range of $J$ values is also observed for the Sbc morphological type, which our Galaxy also belongs to. Two lines with a long and a short dash indicate the behaviour of the median angular momentum given by expressions (\ref{eq:J'}) and (\ref{eq:J_star}), respectively.

As expected, the exclusion of the bulge mass from the angular momentum estimation significantly reduces the median angular momentum for the earliest spirals. Disregarding the gas component slightly reduces the median value of the angular momentum for late-type galaxies. Overall, the general contour of the dependence $J(T)$ remains unchanged.

\begin{figure*}
\centering
\includegraphics[width=\textwidth]{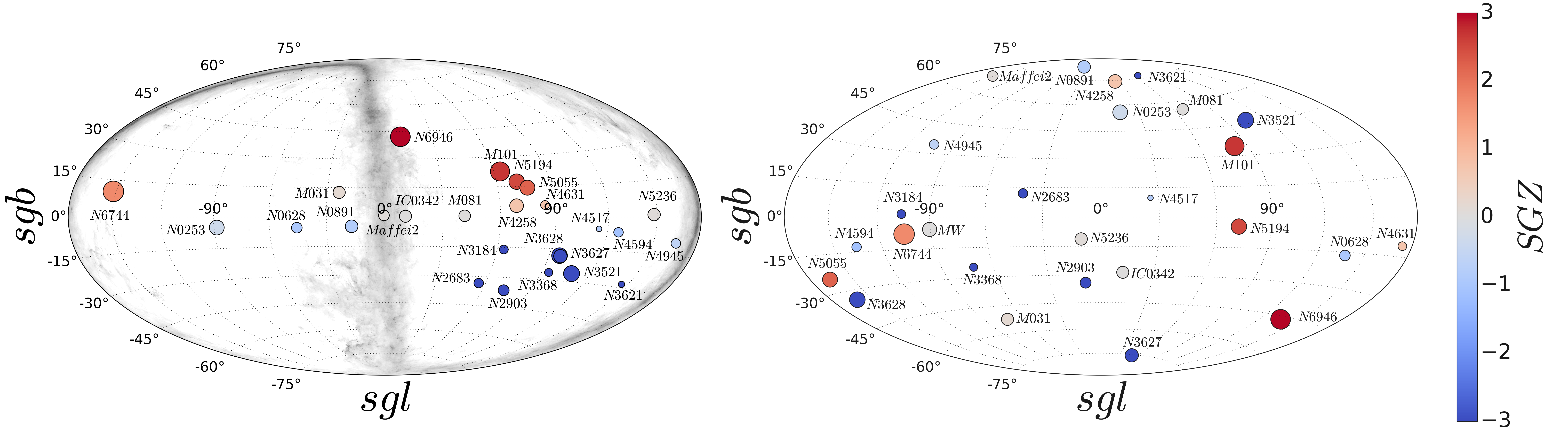}

    \caption{\textit{Left} --- the distribution of galaxies in the Local Volume with large momenta ($J/J_{MW} > 0.15$) in supergalactic coordinates.  \textit{Right} --- the distribution of the spins of the same galaxies in supergalactic coordinates. The size of the circle on both panels characterizes the magnitude of the galactic angular momentum, and its colour characterizes the position of the galaxy relative to the plane of the Local Sheet in Mpc.}
    \label{fig:supergal}
\end{figure*}

\section{Orientation of the spins in the Local Volume}

The direction of the angular momentum of a galaxy, $J$, on the celestial sphere is characterized by equatorial coordinates $\{A,D\}$. If the center of the galaxy has equatorial coordinates $\{\alpha, \delta\}$, the angle between the vector $J$ and the line of sight is $I$, and the position angle of the spin projection onto the picture plane is $\Omega$, these coordinates $\{A,D\}$ can be found from the spherical triangle formulas: 

\begin{equation}
 \begin{array}{ll}
\sin D &= \sin I\times \cos\delta\times \cos\Omega + \cos I\times \sin\delta,\\
\sin(A-\alpha) &= \sin I\times \sin\Omega/\cos D,\\
\cos(A-\alpha) &= (\cos I-\sin\delta\times\sin D)/(\cos\delta\times \cos D).
\end{array} 
\label{eq:DA}
\end{equation}

Here the inclination angle  ranges 
$[0^{\circ}\leq I\leq 180^{\circ}]$, and the position angle with the direction of reference from the declination circle counterclockwise has the range $[-180^{\circ}\leq\Omega\leq 180^{\circ}]$.

The values of the desired angles $I, \Omega$ are usually determined in two steps. At first, we find the position angle of the galaxy's major axis $\omega$ in the range $[0, 180^{\circ}]$ and the inclination of the galaxy's rotation axis to the line of sight in various articles where the kinematics of galaxies was studied. At the second stage, we resolve the ambiguous relations $\Omega=\omega\pm 90^{\circ}$ and $I=\{i, 180^{\circ}-i\}$ using data on the velocity field of the galaxy and its structural features. “The right-hand rule” usually determines the direction of the spin of the galaxy. Thus, if a galaxy is observed edge-on ($i=90^{\circ}$), oriented along the declination circle $(\omega=0)$, and its northern side moves away, then the galaxy's spin is directed to the east. If we see a face-on galaxy ($i=0^\circ)$, the spiral pattern is an indication of the spin's direction. The spin points towards the observer $(I=0^\circ)$ in a galaxy with an S-shaped pattern of arms, and away from the observer $(I=180^\circ)$ in a galaxy with Z-shaped pattern. Here we assume that all spiral galaxies have trailing arms, not leading ones.

In case a galaxy has an arbitrary orientation to the line of sight, we fix the position of its receding side using the velocity field. In addition, we determine the position of the side closest to the observer by the pattern of dust lanes, veins, and spots. The combination of kinematic and photometric (structural) data makes it possible to choose a single spin direction out of four possible ones.

\begin{table*}

\caption{Properties of LV galaxies with $J>0.15J_{MW}$}
 \begin{tabular}{lcccccccccccc}
 \hline
      Name &  Type$^a$ &   $r^a$ &  $\log{L_K/L_{\odot}}^a$ &  $V_m$ &  $A_{26}^a$ &  $I$ &   $\Omega$ &   $J / J_{MW}$ & Twist & To us & Ref. \\
 &   &   (Mpc) &   &  (km s$^{-1}$) &  (kpc) & (deg)  &  (deg)  &   &  &  &  
\\
\hline
Milky Way &     4 &   0.00 &         10.92 &        238 &      30 &   &     & $1.71\cdot 10^{14} {}^*$ &    &  & (1)  \\
MESSIER031         &     3 &  0.77 &        10.73 &        260 &     43.45 &                   77 &                  128 & 0.74 &   ---$^d$ &     r  & (2, 3)\\
IC0342             &     6 &  3.28 &        10.60 &        165 &     34.25 &                   29 &                  132 & 0.73 &     s &     r  & (4)\\
NGC4945            &     6 &  3.47 &        10.66 &        174 &     31.36 &                   97 &                  134 & 0.45 &   ---$^d$ &     r  & (5)\\
NGC0253            &     5 &  3.70 &        10.98 &        200 &     38.28 &                  102 &                 $-41$ & 1.09 &     z &     l  & (5)\\
MESSIER081         &     3 &  3.70 &        10.95 &        250 &     31.93 &                   59 &                   60 & 0.67 &     s &     l & (6) \\
NGC5236            &     5 &  4.90 &        10.86 &        190 &     28.06 &                  156 &                 $-45$ & 0.77 &     z &     l & (7, 8)\\
Maffei2            &     4 &  5.73 &        11.21 &        170 &     17.34 &                  108 &                 $-70$ & 0.57 &     z &     l  & (9, 10)\\
NGC3621            &     7 &  6.64 &        10.34 &        150 &     24.80 &                  115 &                   75 & 0.21 &     z &     l  & (6) \\
MESSIER101         &     6 &  6.95 &        10.79 &        215 &     61.41 &                   20 &                  143 & 1.82 &     s &     r & (4, 8)\\
NGC4631            &     7 &  7.35 &        10.49 &        150 &     33.54 &                   90 &                  176 & 0.38 &   --- &     r  & (11)\\
NGC4258            &     4 &  7.66 &        10.92 &        208 &     40.55 &                   74 &                   61 & 0.93 &     s &     l  & (12)\\
NGC6946            &     6 &  7.73 &        10.99 &        190 &     45.50 &                  141 &                 $-31$ & 1.93 &     z &     l  & (4, 13)\\
NGC4517            &     7 &  8.36 &        10.27 &        141 &     27.52 &                   90 &                   $-6$ & 0.15 &   --- &     l  & (14, $a$)\\
NGC5194            &     5 &  8.40 &        10.97 &        210 &     39.75 &                   22 &                 $-97$ & 1.19 &     s &     r  & (15)\\
NGC2903            &     4 &  8.87 &        10.82 &        215 &     32.45 &                   65 &                 $-66$ & 0.59 &     s &     l & (6) \\
NGC5055            &     4 &  9.04 &        11.00 &        205 &     42.41 &                  121 &                 $-168$ & 1.12 &     z &     r  & (6) \\
NGC6744            &     4 &  9.51 &        10.91 &        200 &     60.42 &                   50 &                  106 & 2.11 &     s &     r & (16)\\
NGC4594            &     1 &  9.55 &        11.32 &        357 &     33.32 &                   96 &                  $179^{HL}$ & 0.45 &   ---$^d$ &     r & (17, $a$)\\
NGC2683            &     3 &  9.82 &        10.81 &        207 &     37.51 &                  102 &                  $-49$ & 0.45 &     z &     l  & (18, $a$)\\
NGC0891            &     3 &  9.95 &        10.98 &        216 &     38.60 &                   90 &                  $-68^{HL}$ & 0.80 &   --- &     l & (19, $a$) \\
NGC0628            &     5 & 10.19 &        10.60 &        200 &     35.39 &                  158 &                  116 & 0.56 &     z &     r  & (20)\\
NGC3368            &     3 & 10.42 &        10.83 &        198 &     27.25 &                  127 &                 $-98$ & 0.33 &     z &     r & (21, $a$)\\
NGC3628            &     4 & 10.52 &        10.96 &        217 &     43.84 &                   90 &                 $-167$ & 1.21 &   --- &     r  & (8, 22)\\
NGC3521            &     4 & 10.70 &        11.09 &        232 &     34.92 &                   73 &                   70 & 1.26 &     s &     l & (6) \\
NGC3627            &     4 & 11.12 &        11.08 &        191 &     33.07 &                   61 &                 $-97$ & 0.88 &     s &     r & (23)\\
NGC3184            &     6 & 11.12 &        10.52 &        240 &     24.27 &                  162 &                 $-93$ & 0.38 &     z &     r & (20)\\
\hline

\end{tabular}
\\
\begin{flushleft}
\textbf{References:} ($a$) UNGC \citet{Karachentsev_Makarov_Kaisina_2013},  ($HL$) HyperLEDA \citet{Makarov_Prugniel_Terekhova_Courtois_Vauglin_2014} (1) \citet{Bland-Hawthorn_Gerhard_2016}, (2) \citet{Carignan_Chemin_Huchtmeier_Lockman_2006}, (3) \citet{Nieten_Neininger_Guelin_Ungerechts_Lucas_Berkhuijsen_Beck_Wielebinski_2006}, (4) \citet{Hernandez_Carignan_Amram_Chemin_Daigle_2005}, (5) \citet{Koribalski_2018}, (6) \citet{de_Blok_2008}, (7) \citet{Comte_1981}, (8) \citet{Sofue_Tutui_Honma_Tomita_Takamiya_Koda_Takeda_1999}, (9) \citet{Kuno_Sato_Nakanishi_Hirota_Tosaki_Shioya_Sorai_Nakai_Nishiyama_Vila-Vilaro_2007}, 
(10) \citet{Hurt_Turner_Ho_1996}, (11) \citet{Rand_1994}, (12) \citet{Marasco_Fraternali_2019}, (13) \citet{Fathi_2007}, (14) \citet{Lee_Wang_2022}, (15) \citet{Colombo_2014}, (16) \citet{Ryder_Walsh_Malin_1999}, (17) \citet{Tempel_Tenjes_2006}, (18) \citet{Kuzio_de_Naray_Zagursky_McGaugh_2009}, (19) \citet{Kamphuis_2007}, (20) \citet{Daigle_Carignan_Amram_Hernandez_Chemin_Balkowski_Kennicutt_2006}, (21) \citet{Nowak_Thomas_Erwin_Saglia_Bender_Davies_2010}, (22) \citet{Wu_Martinez_2022}, (23) \citet{Chemin_2003}\\
\textit{Notes:} $^*$ The dimension of MW angular momentum [kpc km s$^{-1}$ $M_\odot$]. $^d$ Determining the direction of spin from the dust structure. 
When $a$ is in the column Ref., we take  $v_{rot}$ from UNGC, taking into account the new angle of inclination. 
 
\end{flushleft}
\label{tab:Table2}
 \end{table*}

Therefore, we determined the spatial orientation of the spins (where possible) using the available observational data on the structure and kinematics of nearby galaxies. We selected an elite sample of 27 spiral galaxies from the list of LV galaxies (including the Milky Way), whose angular momentum in absolute value exceeds 0.15 of the Milky Way's angular momentum. The total angular momentum of these galaxies is more than 90\% of the total angular momentum of all galaxies in the Local Volume. The main data on elite galaxies are presented in Table~\ref{tab:Table2}. Its columns contain (1) --- galaxy name; (2) --- morphological type; (3) --- distance to the galaxy; (4) --- logarithm of K-band luminosity of galaxy in solar units; (5) --- amplitude of rotational velocity; (6) --- linear diameter; (7) --- inclination of the axis of rotation to the line of sight [from the UNGC catalogue]; (8) --- the position angle of the spin
projection onto the picture plane; (9) --- the value of the total angular momentum, taking into account the gas component, in units of $J_{MW}$; (10) --- the direction of twisting of the spiral pattern (S or Z); (11) --- position of the side of the galaxy moving towards us (r --- right, l --- left), determined from the velocity field. (12) --- References. Links to sources of data on the kinematics of galaxies are given in footnotes to the table. The galaxies are ranked by distance from the observer.

The left panel of Fig.~\ref{fig:supergal} shows the distribution of 26 LV galaxies with the highest angular momenta, shown in supergalactic coordinates. Our Galaxy is present in the figure as a gray band, marking the zone of strong interstellar extinction. The circles represent galaxies, their size reflects the magnitude of the angular momentum. The colour of the circles marks the position of the galaxy relative to the plane of the Local Supercluster (in Mpc) according to the colorbar on the right.

The right panel in Fig.~\ref{fig:supergal}  shows the distribution of elite galaxies in the LV in the direction of their spins in supergalactic coordinates. The size and colour of the circles have the same meaning as on the left panel. The position of the Milky Way's spin is labelled MW. 

We have estimated the uncertainty of spin orientation in this diagram. A comparison of the data on  inclination angles of the galaxies in Table.~\ref{tab:Table2} from UNGC and HyperLEDA catalogs gives the mean square difference of the angles $\sigma(i) =6.6^{\circ}$ and $\sigma(i) =7.9^{\circ}$, respectively. Estimates of the position angles of the major axis of the galaxies from different sources have a scatter $\sigma(PA) =5.0^{\circ}$  compared to HyperLEDA.

As seen from the left panel of Fig.~\ref{fig:supergal}, the plane of the Milky Way is almost perpendicular to the plane of the Local Sheet. It  corresponds to the position of the MW spin near the equator of the Local Supercluster in the right panel of the same figure. This disposition agrees remarkably with the theoretical expectations \citep{Peebles_1969}. However, the nearest massive spiral galaxy M031 has a spin directed 50$^{\circ}$ away from the Local Sheet plane. The spins of other nearby massive spirals (M81, NGC253, IC342) also do not show a tendency to concentrate towards the equator of the Local Supercluster.

\begin{figure}
\centering
\includegraphics[width=\columnwidth]{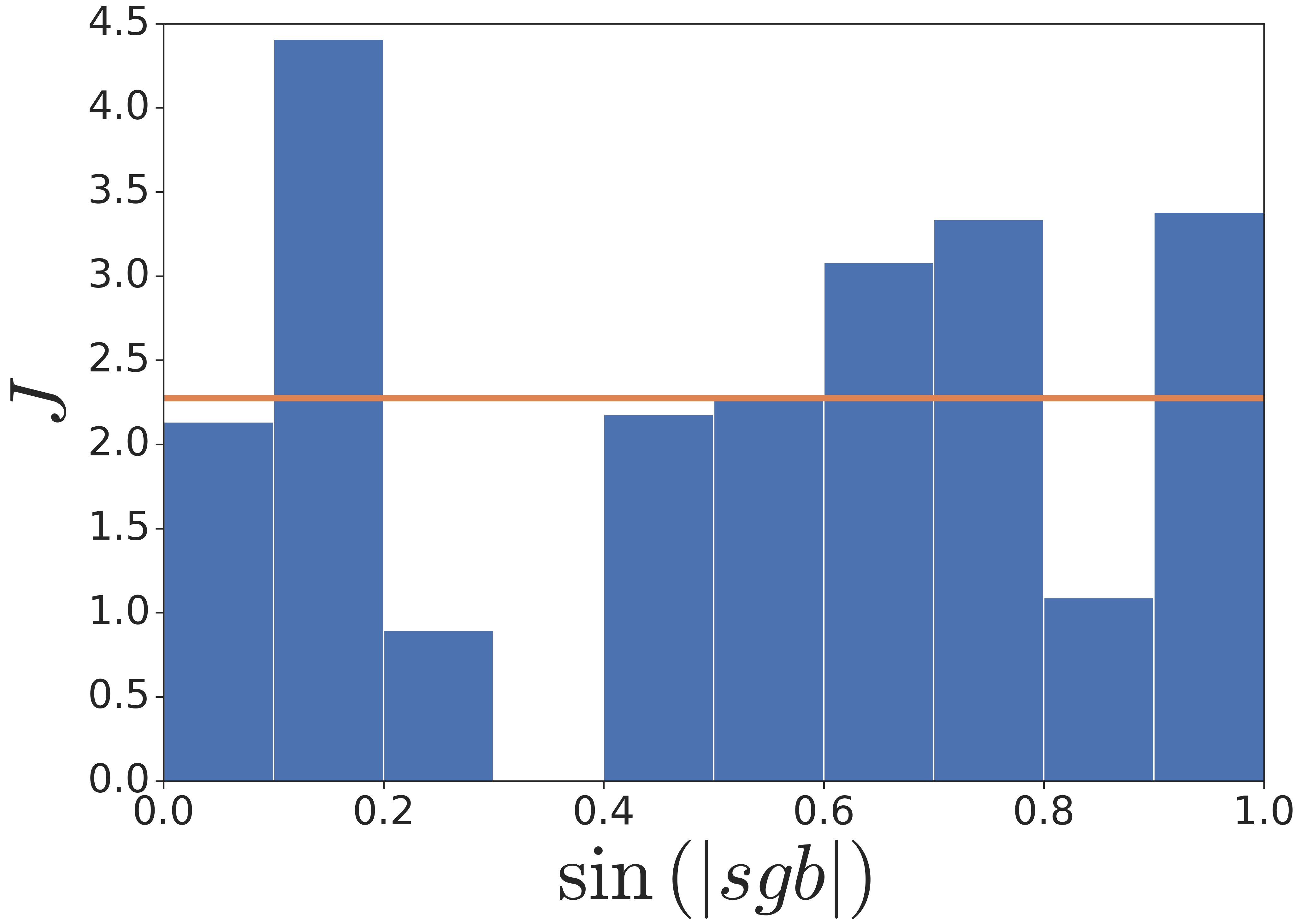}
    \caption{The distribution of 27 spiral galaxies with the largest angular momentum by the sine of the modulus of supergalactic latitude.}
    \label{fig:hist_sinsgb}
\end{figure} 

\begin{figure}

\includegraphics[width=\columnwidth]{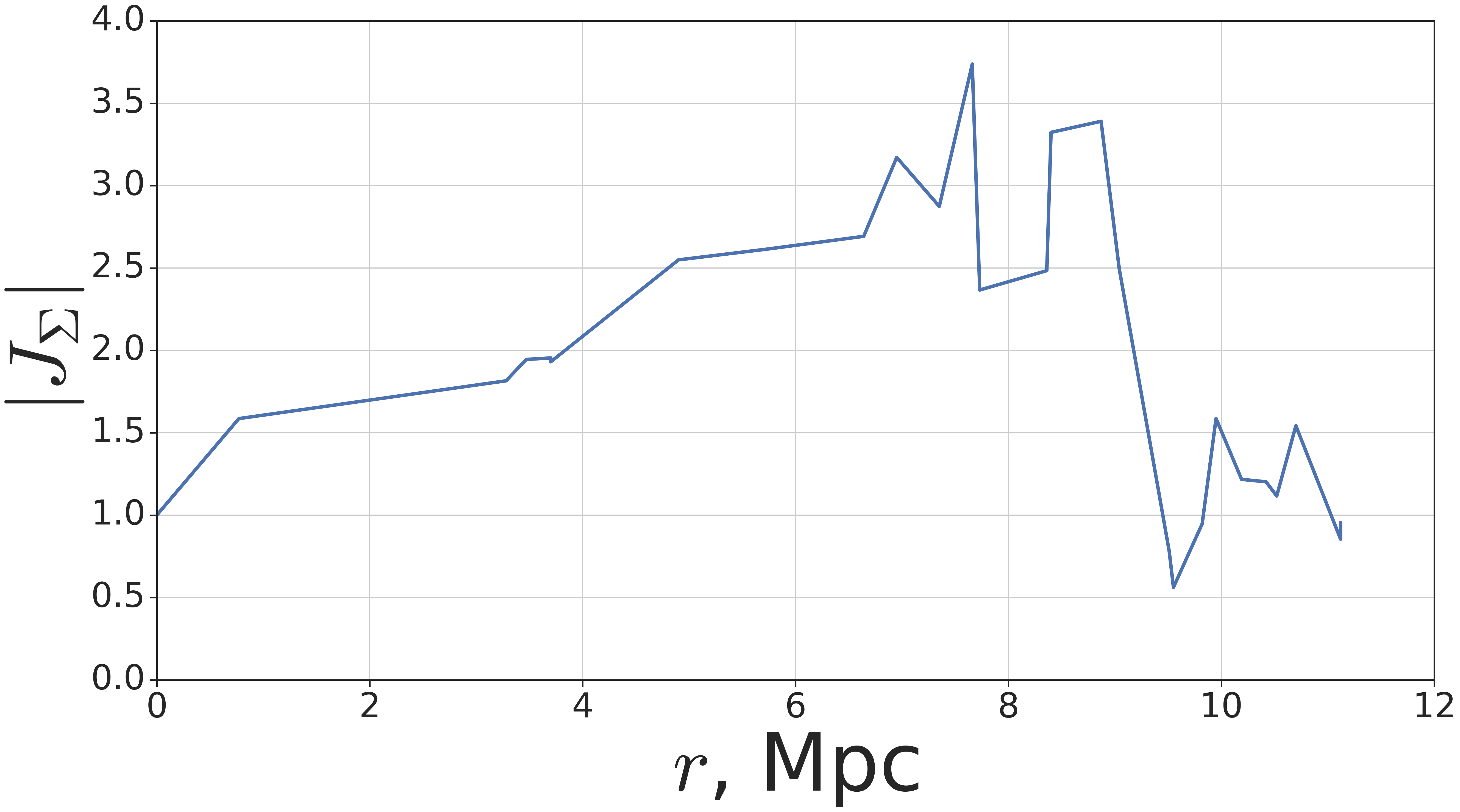}
    \caption{The dependence of the total spin of the Local Volume galaxies on the radius of the sphere they are located in.}
    \label{fig:sum_J}

\includegraphics[width=\columnwidth]{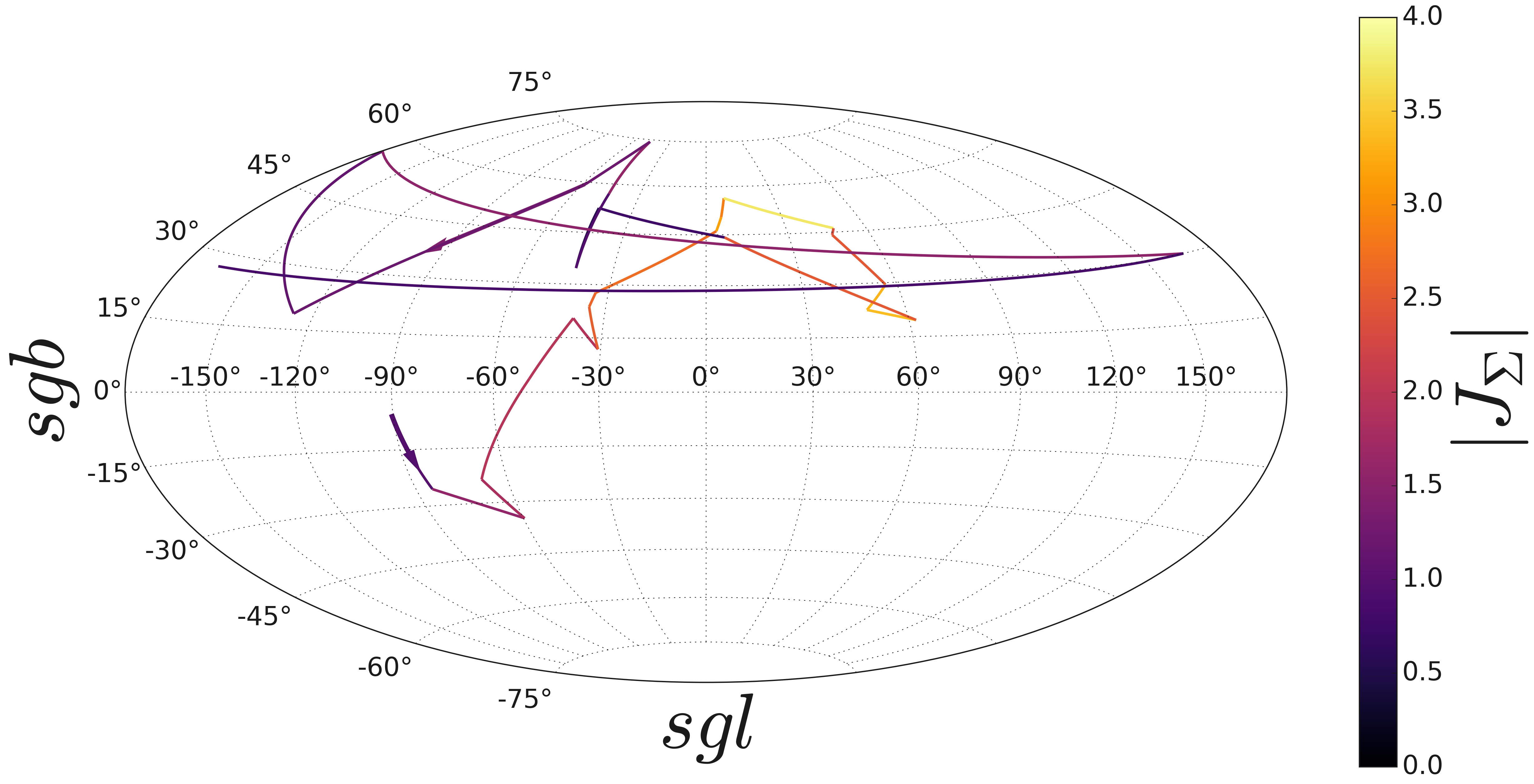}
    \caption{The direction of the total spin of the LV galaxies in supergalactic coordinates as more and more distant galaxies are included in the sample. The colour of the segments on the broken line of “wandering” corresponds to the value of the total moment shown on the right scale.}
    \label{fig:iter}
\end{figure}

The distribution of all 27 LV galaxies with large angular momenta in terms of $\sin |sgb|$ shown in Fig.~\ref{fig:hist_sinsgb}. Its difference from the uniform distribution is not significant according to the $\chi^2$ criterion at the level $P = 0.63$.

Based on the presented data, the orientation of the spins of massive galaxies in the LV looks chaotic and not related to the Local Sheet plane. However, there may be hidden nuances caused by small substructures. For example, \citet{Muller_Scalera_Binggeli_Jerjen_2017a} and \citet{Karachentsev_Neyer_Spani_Zilch_2020}  noted that the massive spiral galaxies NGC 5055, NGC 5194, M 101 (and possibly NGC 6503) form a filament along which distances and radial velocities of the galaxies change smoothly. This chain of groups invades the near part of the Local Void and is oriented at a large angle to the plane of the Local Sheet. The presence of similar small elongated structures: the Sculptor filament (NGC 55, NGC 300, NGC 253, NGC 247, NGC 45) and the M 81 chain (IC 342, NGC 2403, M 81, M 82, NGC 4236), that we see from different angles, can give the impression of a chaotic orientation of the spins.

\begin{figure}
\centering
\includegraphics[width=\columnwidth]{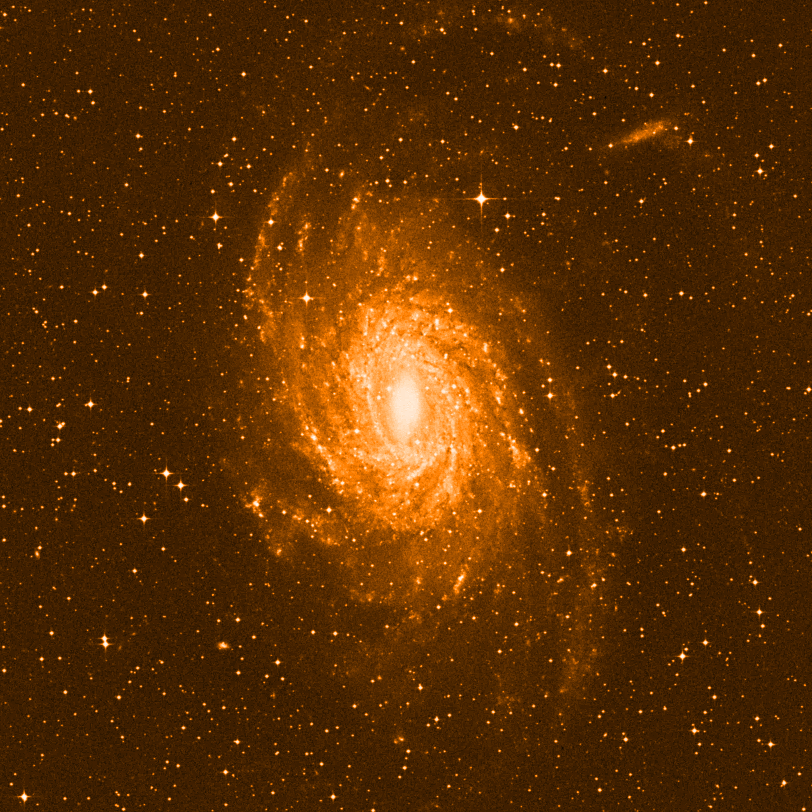}
    \caption{The reproduction of the giant spiral galaxy NGC6744 from the digital Palomar Sky Survey.}
    \label{fig:n6744}
\end{figure}

 We made an attempt to look at the orientation of the angular momentum through the prism of the distance $r$, within which the galaxies are considered. Fig.~\ref{fig:sum_J} shows how the absolute value of the vector sum of the galactic spins changes depending on the distance $r$. The broken-line starts from the accepted value $J_{MW}=1$ for our Galaxy. Each break in the line corresponds to the contribution of the next galaxy to the vector sum of spins. The modulus of the total spin increases from $1$ to $\approx3.5$ within $8.5$ Mpc. $15$ elite spiral galaxies have distances less than it. Then, the total angular momentum falls sharply at $r=9.5$ Mpc due to the contribution of the giant spiral galaxy NGC6744. It is located near the equator of the Local Supercluster (sgb$\approx 10^{\circ}$), and its spin lies almost in the plane of the Local Supercluster (sgb$\approx-7^{\circ}$).  The galaxy NGC6744 is the most extended stellar system in LV with a diameter of $A_{26}=60$ kpc and an angular momentum of about 2 times $J_{MW}$. The dimensions of the galaxy reach 100 kpc in the neutral hydrogen line \citep{Ryder_Walsh_Malin_1999}. A reproduction of this galaxy is shown in Fig.~\ref{fig:n6744}.

\begin{figure}
\centering
\includegraphics[width=\columnwidth]{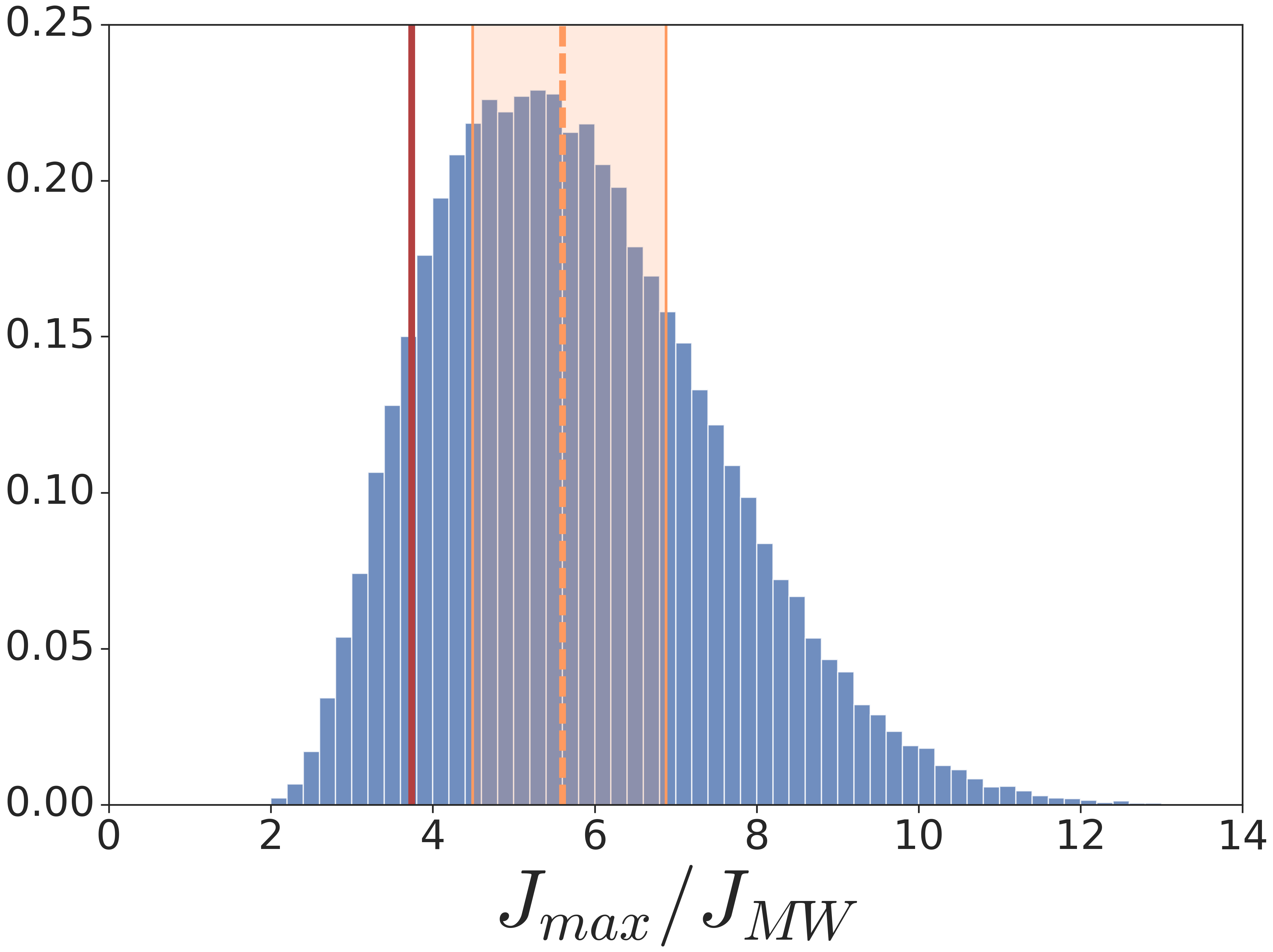}
    \caption{Distribution of the maximum value of the total angular momentum $J_{max}$ in the Monte Carlo simulation with fixed positions of the LV galaxies and random spin directions. The red line shows real $J_{max}$ in the Local Volume, the orange dashed line and region mark the median and area between quantiles of $25\%$ and $ 75\%$, respectively.}
    \label{fig:Jmax}
\end{figure}

Figure~\ref{fig:iter} completes the picture of the behaviour of the total angular momentum of galaxies inside a sphere of increasing radius $r$. It reproduces the wandering of the direction of the total spin of the galaxies on the celestial sphere. The breaks in the “Brownian” track correspond to the contribution of each galaxy as the radius of the sphere increases. The colour of the line reflects the absolute value of the total spin according to the scale shown on the right. As can be seen, the vector of the total spin of galaxies in the LV wanders over the entire celestial sphere. Wherein, at the maximum value of the modulus of the total spin of galaxies (yellow colour in Fig.~\ref{fig:iter}), the vector of the sum of angular momenta remains far from the Local Supercluster plane. 

The question may arise: is the observed peak in Fig.~\ref{fig:sum_J} in the region $r\approx8$~Mpc an indicator of some colinearity of the spins of large galaxies in the LV? To test this assumption, we undertook a series of 100 000 Monte Carlo simulations, fixing the observed distances and magnitudes of angular momenta of the galaxies, but arbitrarily changing their orientation on the celestial sphere. The results are presented in Fig.~\ref{fig:Jmax}, which shows the distribution of the relative number of random realizations over the maximum value of the total angular momentum of 27 galaxies. The observed value
$J_{max}= 3.7$, indicated by the red line, lies not far from the median, 5.6, slightly away from the interquartile band marked by orange. Consequently, the observed data on the spatial orientation of the spins of nearby spiral galaxies do not contain statistically significant indications of their alignment signal.

\begin{figure}
\centering
\includegraphics[width=\columnwidth]{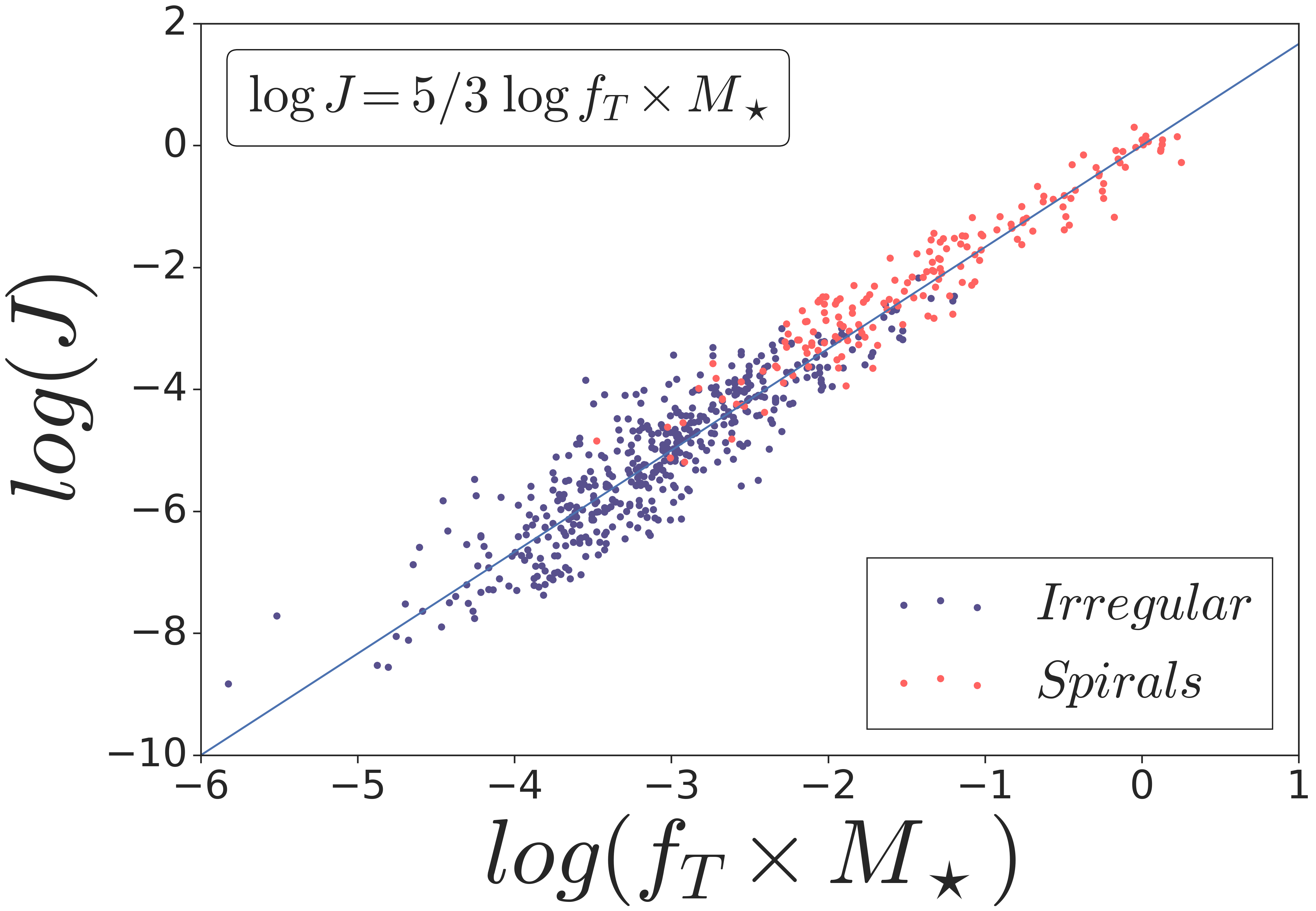}
    \caption{The relationship between an angular momentum and the stellar mass of the disk for spiral and irregular galaxies in the Local Volume. Values $J$ and $f_T \times M_\star$ are normalized to the MW angular momentum and the stellar mass of the MW disk, respectively.}
    \label{fig:MvsJ}
\end{figure}

 Finally, we note that the spiral and irregular galaxies of the Local Volume fit well with the general relationship between stellar mass and total angular momentum \citep{Hardwick_Cortese_Obreschkow_Catinella_Cook_2022}. As follows from the data in Fig.~\ref{fig:MvsJ}, the relation
   
   \begin{equation}
   \log J = (5/3)\times \log f_T \times M_\star+const,
   \end{equation}

determined by the condition of stability of the disks of galaxies, describes the behaviour of small and large galaxies in the mass interval of about 6 orders of magnitude or in the range of rotation angular momenta of about 10 orders of magnitude. The dispersion of galaxies in this diagram is quite large, $\sigma(\log J)\approx0.55$ for irregular and $0.39$ for spiral galaxies. However, \citet{Posti_2018} showed that a more refined account of data on the kinematics and structure of disk galaxies can reduce the dispersion by 3–4 times.

As can be seen from the data in Fig.~\ref{fig:MvsJ}, the specific angular momentum per unit stellar mass of a galaxy is slightly higher for dwarf systems than for spiral ones. This feature was noted by \citet{Chowdhury_Chengalur_2017} and \citet{Kurapati_Chengalur_Pustilnik_Kamphuis_2018}. Obviously, this difference can be reduced if the specific angular momentum of a galaxy is related to the total baryonic mass of its disk.
 
\section{Conclusions}

We have estimated the magnitudes of the angular momentum for seven hundred galaxies enclosed in a volume with a radius of 12 Mpc around the Milky Way. Among them, we selected a sample of 27 elite spiral galaxies whose angular momentum is greater than 0.15 that of the Milky Way. Other galaxies contribute less than 10\% to the total angular momentum of the Local Volume population. For example, M33 has an angular momentum 30 times smaller than the Milky Way, although it is the third largest (by mass) galaxy in the Local Group.

In general, the spin distribution on the sky does not show their preferred orientation to the Local Sheet plane, which may be expected in the standard model of the formation of the angular momentum of galaxies due to the tidal influence of neighbours in the local gravitational field.
 We also considered in the first time the behaviour of the total spin of nearby galaxies as a function of the Local Volume sphere radius $r$. The modulus of the vector sum of angular momentum of the galaxies increases smoothly from $1 J_{MW}$ to $\approx 3.7 J_{MW}$, and then sharply decreases due to the contribution of the momentum from NGC 6744's giant spiral. These 27 elite spiral galaxies have a median deviation from the Local Sheet plane of only 1.2 Mpc. However, the magnitude and orientation of their total spin does not provide a statistically significant indication of the alignment of the galaxies relative to the Local Sheet.

The observed drift of the collective angular momentum of the Local Volume population can be compared to results of cosmic N-body simulations, as was done by \citet{Aragon_Calvo_Silk_Neyrinck_2022}, wherein the spins of dark halos of the galaxies must be taken into account. As noted by \citet{Obreja_Buck_Maccio_2022}, the spin orientation of the dark matter halo may not correlate with the size and mass of the galaxy itself.

The available observational data, in principle, make it possible to advance the analysis of the magnitudes and orientations of the angular momenta of rotation of nearby galaxies further beyond the considered distance limit of $r=12$ Mpc. Expansion of the investigated Local Volume to the distance of the Virgo cluster, 17 Mpc, will require targeted efforts.

\section*{Acknowledgements}
We are grateful to Pavel Kroupa and the anonymous referee for their constructive comments that helped to improve the paper.
This work used the revised version of the Local Volume galaxy database that has been updated within the framework of grant 19-12-00145 of the Russian Science Foundation.

{\bf CONFLICT OF INTEREST}

The authors declare no conflict of interest.

\section*{Data Availability}
The data underlying this article are available in the Local Volume galaxy database https://www.sao.ru/lv/lvgdb/. The additional data will be shared on reasonable request to the corresponding author.



\bibliographystyle{mnras}
\bibliography{article} 








\bsp	
\label{lastpage}
\end{document}